\documentclass[twocolumn,10pt,aps,showpacs,natbib,eqsecnum]{revtex4}
\usepackage{mathptmx} 
\usepackage{graphicx,color}
\usepackage{amssymb}

\mathchardef\bigtilde="0365

\def\la{\langle}
\def\ra{\rangle}
\def\s{\sigma}

\begin{document}

\title{Pad\'e-Borel approximation of the continuum limit of strong coupling lattice fields:\\
Two dimensional non-linear $O(N)$ sigma model at $N\ge 3$}
\author{Hirofumi Yamada}\email{yamada.hirofumi@it-chiba.ac.jp}
\affiliation{%
Division of Mathematics and Science, Chiba Institute of Technology, 
\\Shibazono 2-1-1, Narashino, Chiba 275-0023, Japan}
%}

\date{\today}
%\date{}

\begin{abstract}
{
Based on the strong coupling expansion, we reinvestigate two dimensional $O(N)$ sigma model by the use of Pad\'e-Borel approximants.  The conventional strong coupling expansion of the mass square $M$ in momentum space in $\beta=1/g^2$ is inverted to give $\beta$ expanded in $1/M$.  Borel transform of $\beta$ with respect to $M$ is carried out and the result is improved as the rational function by Pad\'e method.  We find the behavior of Pad\'e-Borel transformed bare coupling at $18$th order is consistent for $N\ge 3$ with that of continuum scaling to the four-loop perturbation theory.  We estimate non-perturbative mass gap at $N\ge 3$ and find the agreement with the exact result by Hasenfratz et.al.
}
\end{abstract}


\maketitle

\section{Introduction}
Nearly four decades ago, the quark confinement was shown by Wilson at the strong bare coupling region \cite{wilson}.   For weak coupling, perturbation theory clarified for the Yang-Mills system that bare coupling $g$ tends to vanish as the lattice spacing $a\to 0$ \cite{pgw}.   The motivation of the present work is to attempt to extrapolate the large $a$ behavior of bare coupling to the asymptotically free behavior at weak coupling.   For the purpose, we like to reformulate the strong coupling expansion by changing the primary variable from bare coupling to the lattice spacing itself.

Lattice serves us a suitable regularization, since in lattice field theories the lattice spacing $a$ explicitly appears in the action and enters into the physical quantities.   For instance, the dimensionless correlation length $\xi$ represents a physical length scale divided by $a$.  It is given at strong coupling as a series ${\cal R}$ in $(g^2)^{-1}$ and it determines, in an implicit manner, the $a$ dependence of the bare coupling.  Mutual roles of $g$ and $\xi$ are exchanged by inverting the relation ${\cal R}$.  Thus we address the question whether the small $\xi$ series of $g$ allows us to confirm directly the weak coupling behavior predicted by perturbation theory.

In the present paper, in non-linear $O(N)$ model at two dimensions, we make an attempt to approximate the asymptotic behavior of bare coupling in the continuum limit via its large $a$ expansion.  The model is of interest as a testing ground of our approach, since it enjoyes asymptotic freedom and dynamical mass generation for $N\ge 3$ \cite{polyakov}.  In addition to the large $N$ limit, we also consider the case of finite $N$.
  
As the basic variable, rather than the correlation length in lattice space, we adopt mass $M$ in momentum space   defined by the zero momentum limit of the two-point field correlation.  The lattice spacing $a$ is included in the mass which is rescaled to be dimensionless and then $M$ must vanish in the continuum limit $a\to 0$ (see (\ref{mass})).  Now the strong coupling expansion gives series of $M$ in $\beta= (g^2)^{-1}$.   By inverting the series, we express $\beta$ as a power series in $1/M$, which is equivalent to large $a$ expansion.  As it would be, naive series fails to confirm the continuum behavior of $\beta$.  However, it is nontrivial and interesting to examine, when both Pad\'e and Borel techniques are applied on the series, whether the continuum scaling emerges at finite $N$ or not.

Before proceeding to following sections, we remark the role of Borel transform in our approach.  We use Borel transform as a device of dilation operation around the continuum limit.     The response of scale transformation on $f(M)$ is probed by rescaling $M$ to $\lambda M$ in $f$ and taking the $\lambda\to 0$ limit.   Then, it is said $f$ scales with the exponent $\Delta$ if
\begin{equation}
f(\lambda M)\to\lambda^{-\Delta}f(M),\quad \lambda\to 0.
\label{scale}
\end{equation}
The above criterion of scaling is implemented by introducing $\delta$ defined by
\begin{equation}
\lambda=1-\delta,\quad 0\le \delta\le 1
\end{equation}
and performing expansion to some finite orders in $\delta$ \cite{yam,yam2}.   Suppose the function approaches to $M^{-\Delta}$.  Then, expanding it to $\delta^{L}$ and setting $\delta=1$, one has $(M(1-\delta))^{-\Delta}\to \frac{L!}{\Gamma(1+\Delta)\Gamma(L-\Delta+1)}M^{-\Delta}$.  Further, if we take the limit $M\to \infty, L\to \infty$ with $M/L=\bar M$ fixed, we obtain
\begin{equation}
M^{-\Delta}\to \frac{1}{\Gamma(1+\Delta)}\Big(\frac{M}{L}\Big)^{-\Delta}=\frac{1}{\Gamma(1+\Delta)}\bar M^{-\Delta}.
\end{equation}
That is, the limit $\lambda\to 0$ has transitioned to the limit $\delta\to 1$ with the cut off $L$.  Then the scaling behavior with exponent $\Delta$ manifests itself in the power of $\bar M (=M/L)$.  
Note the universal quantity $\Delta$ is left unchanged.  
On the other hand, when same operation is acted on $f$ in the series form $\sum a_{k}M^{-k}$ valid at large $M$, we have $\bar f=\sum (a_{k}/k!){\bar M}^{-k}$ with a larger convergence radius,  which is just the Borel transform of the original series.  We thus interpret the Borel transform as a realization of scale transformation.  We do not need integrating $\bar f$ back to $f$.  Though the information of $f(M)$ over the whole range of $M$ is not obtained, what we need in lattice field theories is the behavior of $f(M)$ in the neighborhood of $M= 0$.

\section{Description of the model}
On the two-dimensional square lattice, the continuous spin fields $\vec{\s}=(\s_{1},\s_{2},\cdots,\s_{N})$ are set on every sites.  The action of the system is given by
\begin{equation}
S=-\beta\sum_{\bf n}\sum_{\mu=1,2}\vec{\s}_{\bf n}\cdot\vec{\s}_{\bf n+\bf e_{\mu}},
\label{sigmaaction}
\end{equation}
where ${\bf e}_{1}=(1,0),\,{\bf e}_{2}=(0,1)$ and
\begin{equation}
\beta=\frac{1}{g^2}.
\end{equation}
The fields are constrained to satisfy at every sites, $\vec{\s}^2=N$.

The mass variable $M$ defined via the zero momentum limit of the propagator $(\sum_{\bf n}\exp(i {\bf p}\cdot{\bf n}) \la \vec{\s}({\bf 0})\cdot\vec{\s}({\bf n})\ra)^{-1}$ is given by
\begin{equation}
M=\frac{2D\chi}{\mu}
\label{massdef}
\end{equation}
where susceptibility $\chi$ and second moment $\mu$ are, respectively, given by $\chi=\frac{1}{N}\sum_{{\bf n}}\la \vec{\s}({\bf 0})\cdot \vec{\s}({\bf n})\ra$ and $
\mu=\frac{1}{N}\sum_{{\bf n}}{\bf n}^2\la\vec{\s}({\bf 0})\cdot \vec{\s}({\bf n})\ra$.  
$D$ denotes the dimension of lattice space and $D=2$ in the present work.  Let us summarize the continuum limit of the model and large $a$ expansion of $\beta$.

The perturbative renormalization group predicts that, for $N\ge 3$, the correlation length behaves at weak coupling as
\begin{equation}
\xi= C_{\xi} \exp\Big[\frac{2\pi N\beta}{N-2}\Big]\Big(\frac{2\pi N\beta}{N-2}\Big)^{-1/(N-2)}\Big(1+\sum_{k=1}^{\infty}\frac{a_{k}}{\beta^k}\Big),
\label{rg}
\end{equation}
where the multiplied constant $C_{\xi}$ is specified only non-perturbatively.  Hasenfratz et. al. has computed it via thermodynamic Bethe ansatz \cite{hasen}, giving 
\begin{equation}
C_{\xi}=32^{-1/2}\Big(\frac{e^{1-\pi/2}}{8}\Big)^{\frac{1}{N-2}}\Gamma\Big(1+\frac{1}{N-2}\Big).
\end{equation}

The terms $a_{k}\beta^{-k}$ $(k=1,2,3,\cdots)$ in (\ref{rg}) are contributions of $k+2$-loop levels and three- \cite{fal} and four-loop \cite{col} results were computed in the literature.  They are given as
\begin{eqnarray}
a_{1}&=&\frac{1}{N(N-2)}(-0.0490-0.0141N),\nonumber\\
a_{2}&=&\frac{1}{N^2(N-2)^2}\nonumber\\
&\times &(0.0444+0.0216N+0.0045N^2-0.0129N^3).\qquad
\end{eqnarray}
Though three and higher loop contributions disappear in the continuum limit for the bare coupling, we cannot take out the limit because only the series to finite order is at hand.   Hence, we include known three- and four-loop contributions in our analysis.   

It is known that $M^{-1/2}$ has functional form of $\beta$, the same as (\ref{rg}) but with another multiplicative constant, say $C_{M}$.   However, Monte Carlo data \cite{secondmom} showed that the difference is less than a percent at $N=4$.  Since the two constants agree with each other in the large $N$ limit, the difference between $C_{\xi}$ and $C_{M}$ may actually be negligible for all $N\ge 3$.   Thus the estimation of the mass gap  via strong coupling expansion becomes the estimation of $C_{\xi}$ and this is one of the aims of our work.  .

Since the mass $M$ approaches to $\xi^{-2}$ in the continuum limit, the physical mass of dimension $2$ is given by
\begin{equation}
m^2_{phys}=\lim_{a\to 0}M a^{-2}=C_{\xi}^{-2}\Lambda_{L}^2,
\label{mass}
\end{equation}
where $\Lambda_{L}$ is the finite mass scale given by
\begin{equation}
\Lambda_{L}=a^{-2}\exp\Big[\frac{2\pi N\beta}{N-2}\Big]\Big(\frac{2\pi N\beta}{N-2}\Big)^{-1/(N-2)}\Big(1+\sum_{k=1}^{\infty}\frac{a_{k}}{\beta^k}\Big).
\end{equation}
From (\ref{rg}) and $\xi\sim M^{-1/2}$, we have continuum $\beta$ to four-loop order as a function of $M$,
\begin{eqnarray}
\beta&\sim& \frac{N-2}{4\pi N}\log\frac{x}{C_{\xi}^2}+\frac{1}{2\pi N}\log\Big(\frac{1}{2}\log \frac{x}{C_{\xi}^2}\Big)\nonumber\\
& &+\frac{-2\pi N(N-2)a_{1}+\log\big(\frac{1}{2}\log\frac{x}{C_{\xi}^2}\big)}{\pi N(N-2)\log\frac{x}{C_{\xi}^2}}\nonumber\\
& &+\frac{1}{\pi N(N-2)^2(\log\frac{x}{C_{\xi}^2})^2}\times \nonumber\\
& &\Big[4\pi N(N-2)(-a_{1}+N\pi a_{1}^2-2\pi N a_{2})\nonumber\\
& &+2(1+2\pi N(N-2)a_{1})\log\Big(\frac{1}{2}\log\frac{x}{C_{\xi}^2}\Big)\nonumber\\
& &-\log\Big(\frac{1}{2}\log\frac{x}{C_{\xi}^2}\Big)^2\Big],
\label{twoloop2}
\end{eqnarray}
where
\begin{equation}
x=\frac{1}{M}.
\end{equation}

On the series expansion at large $M$, we borrow the result in the work of Butera and Comi \cite{butera} who computed strong coupling series of $\chi$ and $\mu$ to $\beta^{21}$.   Using the result, we have expansion of $M$ in powers of $\beta$,
\begin{eqnarray}
M&=&\frac{1}{\beta}-4+\frac{2(3+2N)}{2+N}\beta
+\frac{2(16+32N+17N^2+2N^3)}{(2+N)^2(4+N)}\beta^3\nonumber\\
& &-\frac{16(-1+N)}{(2+N)^2}\beta^4+ O(\beta^5).
\end{eqnarray}
By inverting the above relation, we have
\begin{eqnarray}
\beta&=&x-4 x^2+\frac{2 (10 N+19) x^3}{N+2}-\frac{8 (14N+25)
   x^4}{N+2}\nonumber\\
& &+\frac{2 \left(338 N^3+2593 N^2+6084 N+4512\right) x^5}{(N+2)^2 (N+4)}+\cdots.\qquad\quad
\label{strong1}
\end{eqnarray}

Based upon the series (\ref{strong1}), we discuss the approximation of the continuum limit by the use of Pad\'e-Borel approximation scheme. We attempt to recover the asymptotically free behavior (\ref{twoloop2}) from (\ref{strong1}) and then estimate $C_{\xi}$.

\section{Large $N$ limit}
Large $N$ limit serves us a good bench mark of our approach.   So we consider that case first and then turn to finite $N$ in the next section.  

In the large $N$ limit, only the one-loop contribution to $\beta$ survives to give
\begin{equation}
\beta\sim \frac{1}{4\pi}\log (x/C_{\xi}^2),\quad C_{\xi}=(32)^{-1/2}=0.17677669\cdots .
\label{scaling2}
\end{equation}
As briefly presented in the introduction,  Borel transform is given by a certain limit of delta expansion \cite{yam2}.   Explicitly, the logarithm is expanded and gives at $\delta=1$ that
$
\log (x/(1-\delta))\to \log x+\sum_{l=1}^{L}\frac{1}{l}
$
to the order $L$.  Then using the asymptotic expansion $\sum_{l=1}^{L}1/l=\log L+\gamma_{E}+0(L^{-1})$ ($\gamma_{E}$ denotes Euler's constant), we have $\log x\to \log (xL)+\gamma_{E}$ in the $L\to \infty$ limit.  Let $x$ be small enough with $\bar x=xL$ kept finite, then the result represents Borel transform of $\log x$.  
Denoting the operation of Borel transform by ${\cal B}$ we thus find ${\cal B}[\log x]=\log \bar x+\gamma_{E}$.  Using abbreviated simbol $\bar\beta={\cal B}[\beta]$, we then obtain
\begin{equation}
\bar \beta\sim \frac{1}{4\pi}(\log (\bar x/C_{\xi}^2)+\gamma_{E})=\bar\beta_{cont}
\label{oneloop2}
\end{equation}
  
The large $M$ expansion of $\beta$ reads
\begin{equation}
\beta=x-4x^2+20x^3-112x^4+676x^5-4304x^6+\cdots.
\end{equation}
Borel transform of the above series results to divide the $n$th order coefficient by the factorial of $n$,
\begin{equation}
\bar \beta=\bar x-\frac{4}{2!}\bar x^2+\frac{20}{3!}\bar x^3-\frac{112}{4!}\bar x^4+\frac{676}{5!}\bar x^5-\frac{4304}{6!}\bar x^6+\cdots.
\end{equation}
Then as a crucial step, we use Pad\'e method to extrapolate the above series to larger $\bar x$.  The resultant Pad\'e-Borel approximants enable us to capture the scaling behavior to be seen in the scaling region as we can see below.  

As a preliminary study, we have examined the behaviors of $[m/n]$ approximants of $\bar \beta$ over almost possible pairs of $m,n$ at orders $m+n=4,5,\cdots,20$.  On the contrary to the condensed matter models undergoing second order phase transition, critical behavior of the present model is known from perturbation theory as logarithmic and slowly varying.  Hence it is conceivable that 
good behaviors come from the cases where the difference between $m$ and $n$ is small.
The numerical experiment confirmed it is indeed the case.  We have also compared the approximants of three types, Pad\'e-Borel, Borel only and Pad\'e only improvements.  The result at $6$th order is shown in FIG. 1.
\begin{figure}[h]
\centering
\includegraphics[scale=0.75]{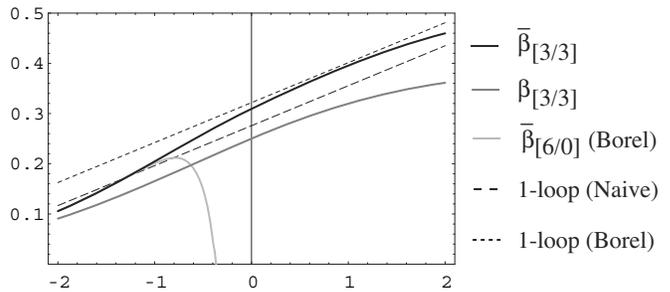}
\label{comparison}
\caption{Plot of improved $\beta$ and $\bar\beta$ at 6th order.   Two broken lines (one for $\beta$ and the other for $\bar \beta$) represent behaviors at continuum.  Horizontal axis corresponds to $\log \bar x=\log(1/\bar M)$ and $\log x=\log(1/M)$ (for Pad\'e only case).}
\end{figure}
As already reported in \cite{yam2}, Borel only improvement is not sufficient for observing the asymptotic freedom.  Pad\'e only case ($\beta_{[3/3]}$ in FIG. 1) is also insufficient as is clear from FIG. 1.  However,  Pad\'e-Borel approximant shows enough improvement for quantitative approximation.  Though Pade only approximation is found to be improved at higher orders, the best performance is achieved by Pad\'e-Borel approximant at every order we analyzed.   We therefore focus on Pad\'e-Borel approximant hereafter.

Now, let us turn to the evaluation of the mass gap by estimating $C_{\xi}$.  Since we know information at weak coupling, the estimation is carried out by fitting $\bar\beta_{cont}$ to $[m/n]$ order approximants of $\bar\beta$, $\bar\beta_{[m/n]}$, by adjusting the value of $C_{\xi}$.  In practice, we consider the difference between $\bar\beta_{[m/n]}$ and $\bar\beta_{cont}$ and plot the difference by changing the value of $C_{\xi}$.  At just proper value of $C_{\xi}$, the two functions touch with each other at a point $\bar x_{0}$ and the difference is tiny over an interval including $\bar x_{0}$.  A typical case is shown in FIG. 2 and the result of estimation of $C_{\xi}$ is shown in Table~\ref{tab:estimation1}.  
For the reason previously written, we list only the results around the diagonal Pad\'e.
\begin{figure}[h]
\centering
\includegraphics[scale=0.75]{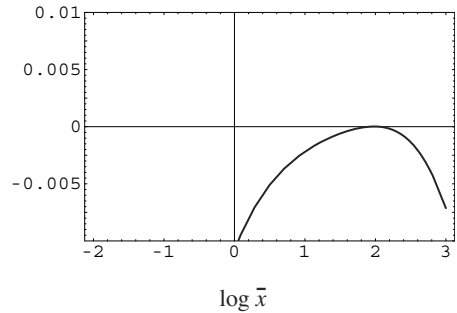}
\caption{Subtracted function $\bar \beta_{[9/9]}-\bar\beta_{cont}=\bar \beta_{[9/9]}-\{\frac{1}{4\pi}(\log \bar x/C_{\xi}^2+\gamma_{E})\}$.  Plotted curve is for $C_{\xi}=0.17868$}
\end{figure}
\begin{table}[h]
\caption{\label{tab:estimation1}
Evaluation result denoted as $C_{app}$ of the non-perturbative constant $C_{\xi}$ in $\frac{1}{4\pi}(\log(\bar x/C_{\xi}^2)+\gamma_{E})$.  Rigorous value of $C_{\xi}$ is $(32)^{-1/2}=0.17677669\cdots$.
 }
\begin{center}
%\begin{tabular}{rl}
\begin{tabular}{cccc}
$  [m/n] $ & $\quad C_{app}$ & $\quad [m/n] $ & $\quad C_{app} $ \\
\hline
$  [3/3]$ & $\quad  0.18327 $  & $\quad [7/7]$ & $\quad  0.17911    $ \\ 
$  [4/3]$ & $\quad  0.18734  $    & $\quad [8/7]$ & $\quad  0.18041     $\\ 
$ [3/4]$ & $\quad  0.18722   $   & $\quad [7/8]$ & $\quad  0.18038     $\\ 
$ [4/4]$ & $\quad  0.18463   $    & $\quad [8/8]$ & $\quad  0.17972     $\\ 
$ [5/5]$ & $\quad  0.18138     $   & $\quad [9/9]$ & $\quad  0.17868    $  \\ 
$ [6/5]$ & $\quad  0.18266     $   & $\quad [10/9]$ & $\quad  0.17900     $  \\ 
$ [5/6]$ & $\quad  0.18264    $   & $\quad [9/10]$ & $\quad  0.17901     $\\ 
$ [6/6]$ & $\quad  0.18178     $   & $\quad [10/10]$ & $\quad  0.17875     $\\ 
\end{tabular}   
\end{center}
\end{table}
Though the reason is not known to us, the orders $6$, $10$, $14$ and $18$ give the best approximation among nearby orders.

\section{Finite $N$ down to $N=3$}
In this section we study the weak coupling behavior from Pad\'e-Borel approximants for a finite number of spin components.   First we discuss Borel transform of (\ref{twoloop2}) to compare it with Pad\'e-Borel approximants of large $M$ series (\ref{strong1}). 

 Let us consider Borel transform of the two-loop contribution.  We find
\begin{eqnarray}
{\cal B}[\log\log x]=\log\log \bar x+\frac{\gamma_{E}}{\log \bar x}+\frac{\zeta(2)-\gamma_{E}^2}{2(\log \bar x)^2}+O((\log \bar x)^{-3}).\quad
\label{borel}
\end{eqnarray}
The result is simplified by absorbing $\gamma_{E}$ into the log.  Then we obtain
\begin{equation}
{\cal B}[\log\log x]\sim  \log(\log \bar x+\gamma_{E})+\frac{\zeta(2)}{2(\log \bar x+\gamma_{E})^2}+O((\log \bar x)^{-3}).
\end{equation}
Note that the second term should be included when the four loop contribution is taken into account.  At two- and three-loop orders, we need only the first term.  In a similar manner, for contributions at three- and four-loop orders, we find as follows:
\begin{eqnarray}
{\cal B}[\frac{1}{\log x}]&=&\frac{1}{\log \bar x+\gamma_{E}}+O((\log \bar x)^{-3}),\\
{\cal B}[\frac{\log\log x}{\log x}]&=&\frac{\log(\log \bar x+\gamma_{E})}{\log \bar x+\gamma_{E}}+O((\log \bar x)^{-3}),\\
{\cal B}[\frac{1}{(\log x)^2}]&=&\frac{1}{(\log \bar x+\gamma_{E})^2}+O((\log \bar x)^{-3}),\\
{\cal B}[\frac{\log\log x}{(\log x)^2}]&=&\frac{\log(\log \bar x+\gamma_{E})}{(\log \bar x+\gamma_{E})^2}+O((\log \bar x)^{-3}),\\
{\cal B}[\frac{(\log\log x)^2}{(\log x)^2}]&=&\frac{(\log(\log \bar x+\gamma_{E}))^2}{(\log \bar x+\gamma_{E})^2}+O((\log \bar x)^{-3}).\qquad 
\end{eqnarray}

Thus the result of Borel transform to the four-loop level reads,
\begin{eqnarray}
\bar \beta&=&\frac{N-2}{4\pi N}(\log \frac{\bar x}{C_{\xi}^2}+\gamma_{E})+\frac{1}{2\pi N}\log[\frac{1}{2}(\log \frac{\bar x}{C_{\xi}^2}+\gamma_{E})]\nonumber\\
& &+\frac{-2\pi N(N-2)a_{1}+\log[\frac{1}{2}(\log \frac{\bar x}{C_{\xi}^2}+\gamma_{E})]}{\pi N(N-2)(\log\frac{\bar x}{C_{\xi}^2}+\gamma_{E})}\nonumber\\
& &+\frac{1}{\pi N(N-2)^2(\log\frac{\bar x}{C_{\xi}^2}+\gamma_{E})^2}\times \nonumber\\
& &\Big[4\pi N(N-2)(-a_{1}+N\pi a_{1}^2-2\pi N a_{2})\nonumber\\
& &+2(1+2\pi N(N-2)a_{1})\log[\frac{1}{2}(\log\frac{\bar x}{C_{\xi}^2}+\gamma_{E})]\nonumber\\
& &-\log[\frac{1}{2}(\log\frac{\bar x}{C_{\xi}^2}+\gamma_{E})]^2\Big]+\frac{\zeta(2)}{4\pi N(\log\frac{\bar x}{C_{\xi}^2}+\gamma_{E})^2}\nonumber\\
&=&\bar\beta_{cont}.
\label{twoloop3}
\end{eqnarray}

\begin{figure}[h]
\centering
\includegraphics[scale=0.6]{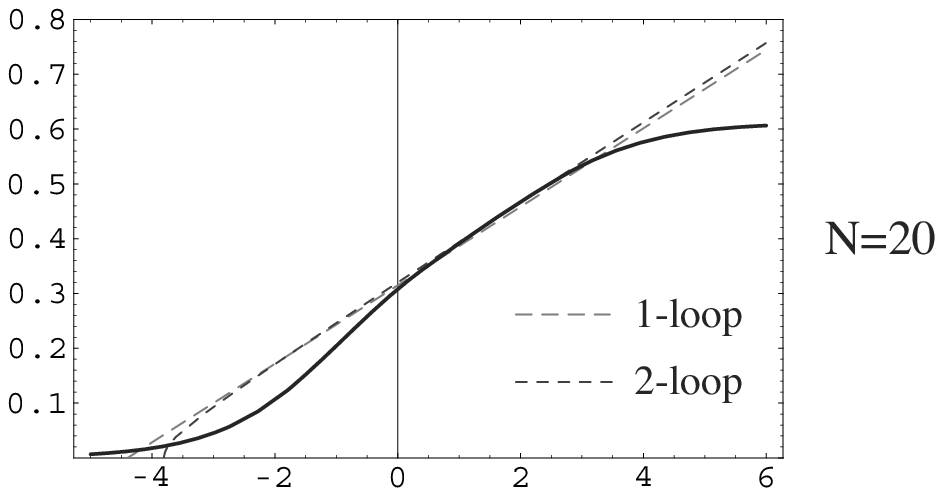}
\includegraphics[scale=0.6]{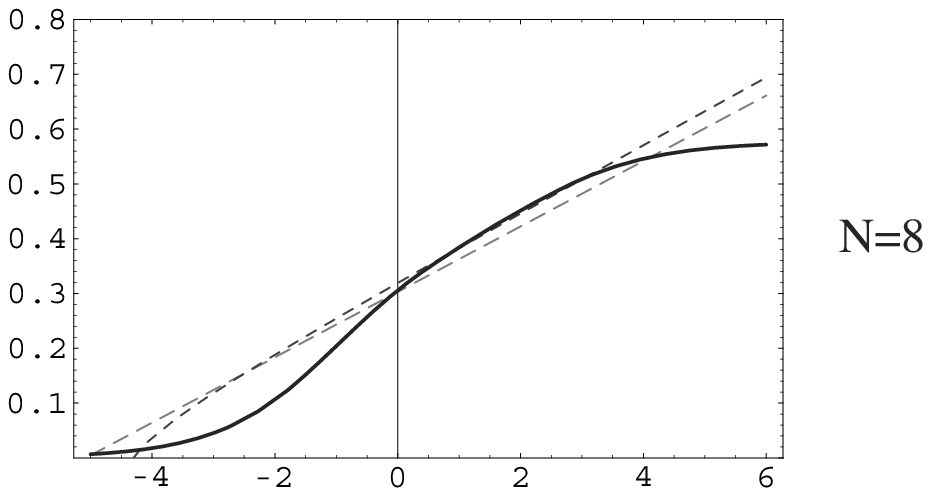}
\includegraphics[scale=0.6]{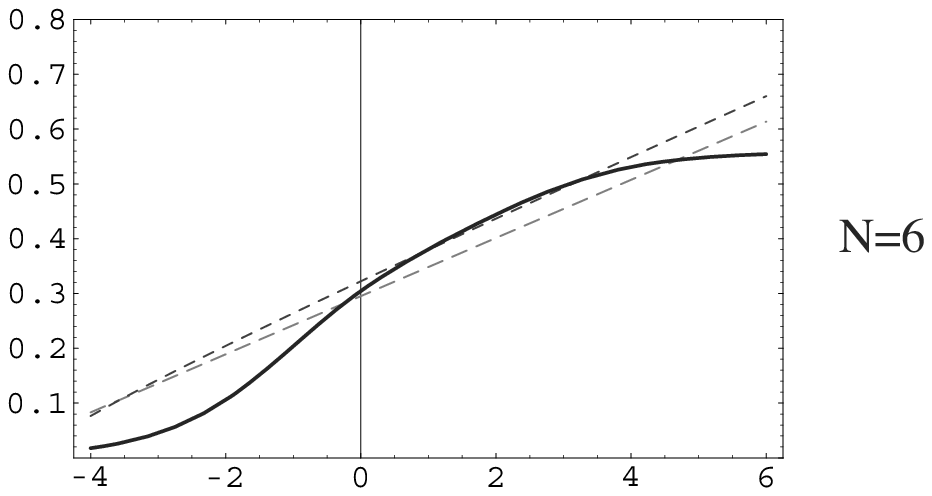}
\includegraphics[scale=0.6]{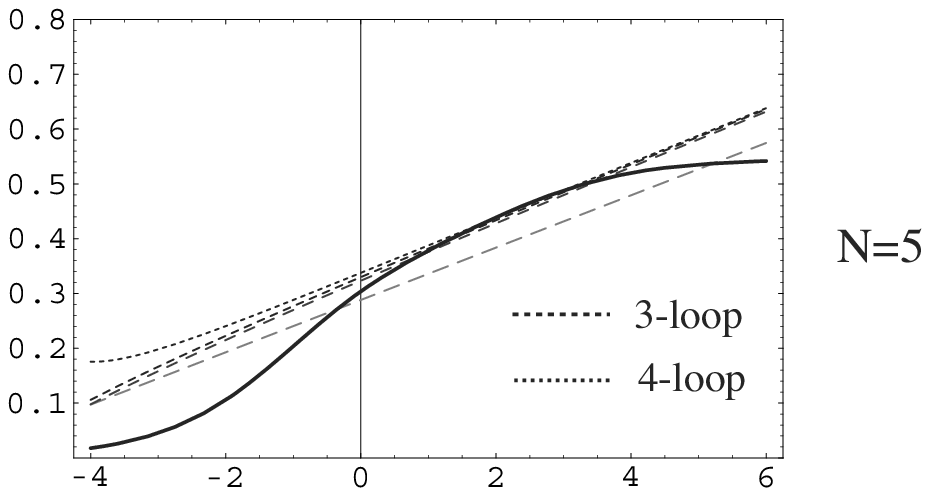}
\includegraphics[scale=0.6]{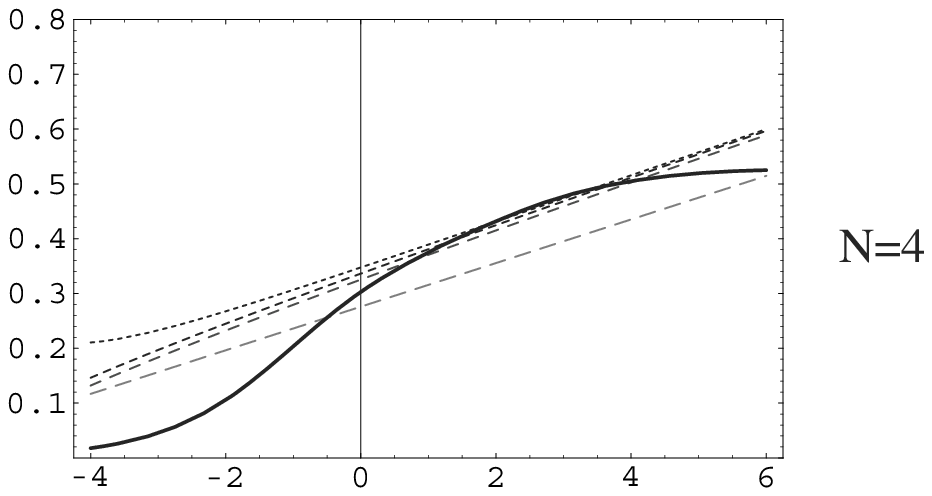}
\includegraphics[scale=0.6]{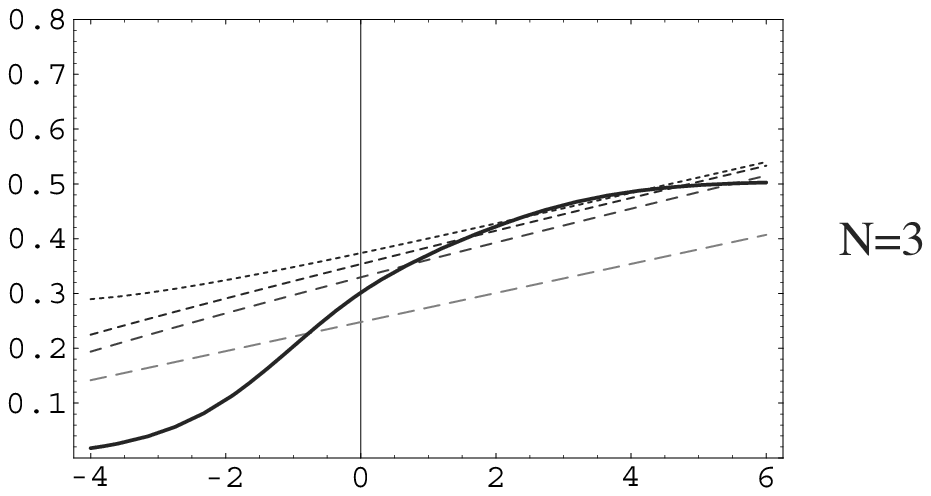}
\caption{Plots of  $\bar\beta_{[9/9]}$ at $N=20, 8, 6, 5, 4, 3$ and $\bar\beta_{cont}$ at one and two-loop results (plus three and four-loop results for $N=3, 4$ and $5$) as functions of $\log\bar x$.}
\end{figure}

At large $\bar M$, we have from (\ref{strong1}),
\begin{equation}
\bar \beta=\bar x-\frac{4}{2!} \bar x^2+\frac{2 (10 N+19) \bar x^3}{3!(N+2)}-\frac{8 (14N+25)
   \bar x^4}{4!(N+2)}+\cdots.
\label{strongborel}
\end{equation}
As in the previous section, we further improve the large $\bar M$ series by Pad\'e method.  We have checked that also at finite $N$, diagonal Pad\'e provides best behaviors.  Skipping low order results, we explicitly present only the results at $18$th order for various $N$.  FIG. 3 shows the plots of $\bar\beta_{[9/9]}$ and $\bar\beta_{cont}$ at one- and two-loop levels (at $N=3,4,5$, $\bar\beta_{cont}$ at three- and four-loop levels are also plotted) as  functions of $\log\bar x$.  At $N\ge 6$, three- and four-loop $\bar\beta_{cont}$ are very close to that at two-loop at $\bar x>0$ and we have omitted them.
At $N=3, 4$, though the scaling to the four-loop level is not so clear, the behavior of $\bar\beta_{[9/9]}$ is roughly consistent with the continuum one for $\log\bar x \in[2,4]$.  At $N=5$, linear-like behavior with correct slope is observed around $\log\bar x\sim 2$, which signals scaling behavior.  From $N\sim 8$, we observe continuum scaling at the two-loop level.  
  
Now, having examined continuum scaling, we evaluate constant $C_{\xi}$ as in the same manner at $N=\infty$.  Namely we consider $\bar\beta_{[9/9]}-\bar\beta_{cont}$ and search for the value of $C_{\xi}$ by fitting $\bar\beta_{cont}$ to $\bar\beta_{[9/9]}$ by changing values of $C_{\xi}$.   
\begin{table}[h]
\caption{\label{tab:estimation2}
Result of estimation of the constant $C_{\xi}$ (implied by $C_{app}$).  The last column shows the result of Botera and Comi \cite{butera}.}
\begin{center}
%\begin{tabular}{rl}
\begin{tabular}{clllll}
$N$ & $\quad C_{app}^{2-loop}$ & $\quad C_{app}^{3-loop}$ & $\quad C_{app}^{4-loop}$  & $ \quad C_{\xi}$ & $\quad C_{BC}$  \\
\hline
$  3$ & $\quad  0.0068    $  &  $\quad 0.0094$ &  $\quad 0.0112$ &   $\quad 0.0125$   &  \\ 
$  4$ & $\quad  0.0336    $  &  $\quad 0.0373$ &   $\quad 0.0398$ &   $\quad 0.0416$   &  $\quad 0.039$ \\ 
$  5$ & $\quad  0.0584   $    &  $\quad 0.0615$ &  $\quad 0.0639$ & $\quad 0.0652$  &  $\quad 0.065$\\ 
$6$ & $\quad  0.0771     $   & $\quad 0.0797$ &  $\quad 0.0818$ & $\quad 0.0826$  &  $\quad 0.084$\\ 
$ 7$ & $\quad  0.0913      $  & $\quad 0.0934$ &  $\quad 0.0953$ & $\quad 0.0955$   &  \\ 
$ 8$ & $\quad  0.1021    $   & $\quad 0.1038$ &  $\quad 0.1055$  &  $\quad 0.1054$  &  $\quad 0.106$\\ 
$ 9$ & $\quad  0.1106     $  & $\quad 0.1121$ &   $\quad 0.1136$  & $\quad 0.1132$  &  \\ 
$ 10$ & $\quad  0.1175    $  &  $\quad 0.1187$ &     $\quad 0.1201$  &   $\quad 0.1195$  &  $\quad 0.121$\\ 
$ 11$ & $\quad  0.1231    $   & $\quad 0.1242$ &   $\quad 0.1255$  & $\quad 0.1247$  &  \\ 
$ 12$ & $\quad  0.1278    $   &  $\quad 0.1288$ &  $\quad 0.1299$  & $\quad 0.1290$  &  $\quad 0.130$\\ 
$ 13$ & $\quad  0.1318    $   &  $\quad  0.1326$ &   $\quad 0.1337$  &  $\quad 0.1327$  &  \\ 
$ 14$ & $\quad  0.1352    $   &  $\quad 0.1359$ &  $\quad 0.1369$  &  $\quad 0.1358$  &  $\quad 0.137$\\ 
$ 15$ & $\quad  0.1381    $   &  $\quad 0.1388$ &  $\quad 0.1397$  & $\quad 0.1386$  &  \\ 
\end{tabular}   
\end{center}
\end{table}
\begin{figure}[h]
\centering
\includegraphics[scale=0.75]{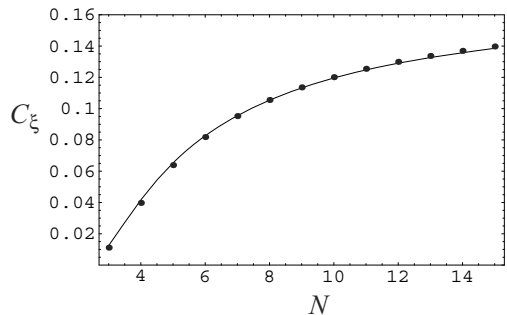}
\caption{Plot of true value of $C_{\xi}$ (solid curve) and its estimation at $N=3, 4,5,6,\cdots, 15$ (black points) carried out via Pad\'e-Borel approximants $\bar\beta_{[9/9]}$ compared with four-loop $\bar\beta_{cont}$.}
\end{figure}
The result is summarized in TABLE II and FIG. 4.  We may say that, for all $N\geq 3$, especially for $N=3$ and $4$, $\bar\beta_{[9/9]}$ yields a good four-loop estimation of non-perturbatice constant $C_{\xi}$.   However we see that, at $N\geq 8$, the estimated value $C_{app}$ is slightly larger than the exact one.  Note that an excess of the estimation is observed also in the large $N$ limit.  On the contrary, for $N\le 7$, $C_{app}$ is slightly smaller than $C_{\xi}$.   We like to discuss the issue in the next section.
  To summarize, we conclude that the approximation level is satisfactory.

\section{Discussion}
From previous two sections, at larger $N$, we found that the four-loop estimation of $C_{\xi}$ gives excess to the exact value.    For example, at $[9/9]$ approximants, the excess reads $\sim 0.0001$, $\sim 0.0011$ and $\sim 0.0019$ at $N=8, 15$ and $\infty$, respectively.  As long as $\bar x$ is not so small, every multiple-loop (above one-loop level) contribution decreases as $N$ becomes large and vanishes at $N=\infty$.   Hence, at large enough $N$, the five, sixth, $\cdots$-loop contributions may be safely neglected in our study.   Then the main factor of the discrepancy would come from the lattice artifact.  To find evidence, let us 
discuss the large $N$ limit since that case provides us quantitative example as we can see below.  

At $N=\infty$,  $\beta$ behaves for small $M$ as
\begin{equation}
\beta=-\frac{1}{4\pi}\log\frac{M}{32}+\frac{M}{32\pi}(\log\frac{M}{32}+1)+O(M^2\log M).
\label{original}
\end{equation}
Note that this is just the one-loop result.  The second and higher order terms represent lattice artifacts which disappear in the continuum limit.  They involve the logarithm
and delay the approach of $\beta$ to the continuum limit.   In fact, use of Borel transform has the notable advantage that it reduces the correction to
\begin{equation}
\bar\beta=-\frac{1}{4\pi}(\log\frac{\bar M}{32}-\gamma_{E})-\frac{\bar M}{32\pi}+O(\bar M^2).
\end{equation}
Here we have used
\begin{eqnarray}
{\cal B}[M\log M]&=&-\bar M,\\
{\cal B}[M]&=&0.
\end{eqnarray}
For original $\beta$ (see (\ref{original})), the second term is of order $M\log M$, but for $\bar\beta$, $\bar M$ and the deviation from the asymptotic scaling is much reduced when $\bar M$ is small enough.  
The correction, however, still affects the small $\bar M$ behavior of the transformed bare coupling.   We have examined scaling and evaluated $C_{\xi}$ by keeping the first order correction $-\frac{\bar M}{32\pi}$.  From TABLE~\ref{tab:estimation3}, it is apparent that incorporation of $O(M\log M)=O(a^2\log a^2)$ term improves the approximation.  At $[9/9]$ order the excess is only $\sim 0.0001$.  This means that Pad\'e-Borel approximants  actually recover the small $M$ behavior very well, but in the same time, the residual effect of the correction is still non-negligible for higher accuracy.   We thus find that main factor of the discrepancy comes from the lattice artifact as long as $N$ is large enough.
\begin{table}[h]
\caption{\label{tab:estimation3}
Result of estimation of the constant $C_{\xi}$ (implied by $C_{app}$) in the large $N$ limit when the correction $-\frac{\bar M}{32\pi}$ to the asymptotic scaling is taken into account.  Only the results of diagonal approximants are shown.   Rigorous value of $C_{\xi}$ is $(32)^{-1/2}=0.17677669\cdots$. }
\begin{center}
%\begin{tabular}{rl}
\begin{tabular}{cl}
$[m/n]$ & $\qquad C_{app}$  \\
\hline
$[2/2]$ & $\qquad  0.187126    $  \\ 
$ [3/3]$ & $\qquad  0.177873    $  \\ 
$[4/4]$ & $\qquad  0.178509   $   \\ 
$[5/5]$ & $\qquad  0.177355     $    \\ 
$ [6/6]$ & $\qquad  0.177484      $   \\ 
$ [7/7]$ & $\qquad  0.176892  $    \\ 
$ [8/8]$ & $\qquad  0.177036     $   \\ 
$ [9/9]$ & $\qquad  0.176891   $     \\ 
$ [10/10]$ & $\qquad  0.176900    $    \\ 
\end{tabular}   
\end{center}
\end{table}

Next, consider the case of lower $N=3\sim 7$ where the estimated value of $C_{\xi}$ is slightly lower than the exact one.  As a typical case, consider the $N=3$ case.  Three- and four-loop effects contribute to $\bar\beta_{cont}\sim  0.4694$ at $x=\bar x_{0}$ ($\log \bar x_{0}\sim 3.2$) by amounts $\sim 0.02$ and $\sim 0.01$, respectively.  Though they carry with small fractions of total $\bar \beta$,  they are not negligible at all, since $C_{app}$ increases by $0.0026$ and $0.0018$ when three- and four-loop effects is taken into account, and the magnitude of $C_{\xi}$ itself is small.  Therefore, loop contributions above four would be still active for estimating $C_{\xi}$ and even have possibility to push $C_{app}$ be larger than $C_{\xi}$.   For small $N$, in addition to the lattice artifact, a discrepancy may come also from lack of higher loops.

On the lattice artifact, it is crucial to reduce it for obtaining precise result for all $N\geq 3$.   It has reported in \cite{balog} that the standard action gives $a^2(\log a^2)^{N/(N-2)}$ as the leading lattice artifact near the continuum limit.  It has the maximum value at $N=3$ giving contribution $\sim a^2(\log a^2)^{3}$.   Borel transform would reduce the effect of such a logarithmic term but the effect would remain to obscure the asymptotic scaling at finite $\bar x$. 

In general, the leading lattice artifact may not be known completely.  Then, one way to resolve the issue is to construct or use lattice action in which such artifacts are reduced from the outset.  As an example, we report the result of Symanzik's modification of lattice action \cite{sym} in the large $N$ limit.   In Symanzik improvement program, one generalizes the action element from $1-\vec{\s}_{\bf n}\cdot\vec{\s}_{\bf n+\bf e_{\mu}}$ to $\sum_{k=0}^{K}A_{k}\vec{\s}_{\bf n}\cdot\vec{\s}_{{\bf n}+k{\bf e_{\mu}}}$.  By expanding the action in $a$ and minimizing lattice artifact at the level of action, one can obtain the optimized set of coefficients $A_{k}$ $(k=0, 1,2,3,\cdots, K)$.  Then, the direct effect is the modification of the unperturbed propagator from $[2\sum_{\mu=1,2}(1-\cos ap_{\mu})]^{-1}=[\sum_{\mu}a^2 p_{\mu}^2-\sum_{\mu}a^4 p_{\mu}^4/12+O(a^6 p^6)]^{-1}$ to the one closer to the continuum limit $[a^2 \sum_{\mu}p^2]^{-1}$.   For instance, to the first order ($K=2$) we have 
\begin{eqnarray}
& &[\sum_{\mu}(\frac{5}{2}-\frac{8}{3}\cos ap_{\mu}+\frac{1}{6}\cos 2ap_{\mu})]^{-1}\nonumber\\
&=&[a^2\sum_{\mu}p^2-a^6\sum_{\mu}p_{\mu}^6/90+O(a^8 p^8)]^{-1}.
\end{eqnarray} 
At the infinite order ($K=\infty$) the action becomes infinite series composed of field couplings between two sites along $\mu$ of all distances.
\begin{table}[h]
\caption{\label{tab:estimation4}
Ratio of $C_{app}$ (approximant of $C_{\xi}$) to the exact value of $C_{\xi}$ in standard, first order and infinite order improved actions.  The blanks represent absence of extremum zero of $\bar\beta_{[n/n]}-\bar\beta$.  However, even in those cases, the difference (the subtracted function) exhibits almost stationary behavior around the point, say also $\bar x_{0}$, at which $\bar\beta_{[n/n]}-\bar\beta$ vanishes and the first derivative takes minimum value.  The estimation of $C_{\xi}$ at such $\bar x_{0}$ yields accurate values.}
\begin{center}
%\begin{tabular}{rl}
\begin{tabular}{clll}
$[m/n]$ & $\quad standard $ & $\quad 1st\,\, order $ & $\quad infinite\,\, order $  \\
\hline
$  [3/3]$ & $\quad  1.03673 $  &   $\quad 1.01102$    &   \\ 
$ [4/4]$ & $\quad  1.04443   $  &   $\quad 1.01137$   & $\quad 0.999244$ \\ 
$ [5/5]$ & $\quad  1.02604     $    &  $\quad 1.00505$  &   \\ 
$ [6/6]$ & $\quad  1.02830    $   &  $\quad 1.00516$  & $\quad 0.999914$\\ 
$ [7/7]$ & $\quad  1.01320    $  &  $\quad 1.00168$    &  \\ 
$ [8/8]$ & $\quad  1.01665    $   &  $\quad 1.00199$  & $\quad 0.999990$\\ 
$ [9/9]$ & $\quad 1.01077   $  &   $\quad 1.00103$  &  \\ 
$ [10/10]$ & $\quad  1.01145 $  &   $\quad 1.00103$  & $\quad 0.999999$\\ 
\end{tabular}   
\end{center}
\end{table}
The result in momentum space is simple modification of the propagator to the continuum limit $[a^2 \sum_{\mu}p^2]^{-1}$.  Since in the large $N$ limit, $\beta$ is given by the gap equation written only with the propagator with mass square $M$, we can easily obtain the large $M$ series both at first- and infinite-order improved actions (For detailed presentation, see the first reference in \cite{yam}).  For example, at the first order it follows
\begin{eqnarray}
\beta&=&\int_{-\pi}^{\pi}\frac{d^2 p}{(2\pi)^2}\frac{1}{M+\sum_{\mu=1,2}(\frac{5}{2}-\frac{8}{3}\cos p_{\mu}+\frac{1}{6}\cos 2p_{\mu})}\nonumber\\
&=&\frac{1}{M}-\frac{5}{M^2}+\frac{1157}{36M^3}-\frac{8419}{36M^4}+O(M^{-5}).
\end{eqnarray}
At infinite-order improvement, the right-hand side becomes just the integral of $(M+\sum_{\mu=1,2}p_{\mu}^2)^{-1}$ and expansion of $\beta$ in $1/M$ is straightforward.  
It now suffices for us to repeat the same procedure for the approximation of (\ref{scaling2}) and the constant $C_{\xi}$ at the first and inifinite orders of improved actions.    Here note that the change of action induces the change of the value of non-universal $C_{\xi}$.   $C_{\xi}=0.2377607\cdots$ and $0.2851456\cdots$ at first and infinite orders, respectively.   TABLE~\ref{tab:estimation4} summarizes the result of our approximation.  The improved action improves the approximation accuracy both at the first and at infinite orders.   Though the improved lattice action is conventionally used in the Monte Carlo analysis and perturbation theory, it is also useful in our approach.  

In the present work, we have analyzed Pad\'e-Borel approximants of strong coupling expansion in non-linear $\sigma$ model and have found good behaviors approximating the continuum limit.   We close the paper by pointing out that, even working with the standard action, further higher order computation would improve the result for all $N$ including the limit $N\to \infty$.   Pad\'e-Borel approximants may become effective at larger $\bar x$ (smaller $\bar M$) and the two unwanted  effects, lattice artifacts and omitted loop contributions, would be weaker there.  Then, continuum scaling at smaller $\bar M$ with a clearer sign of asymptotic freedom near $N=3$ would be seen, which allows us accurate evaluation of the mass gap for all $N\ge 3$.

%\newpage
%\section*{References}

\end{document}